\documentclass[superscriptaddress,twocolumn,aps]{revtex4}

\usepackage{harvard}
\renewcommand\harvardand{and}

\usepackage{amsmath}
\usepackage{amssymb}

\usepackage{graphicx}
\graphicspath{{./Figures/}}

\usepackage{psfrag}
\usepackage{subfigure}

\newcommand\phib{\phi_{\rm{b}}}
\newcommand\xib{\xi_{\rm{b}}}
\newcommand\xif{\xi_{\rm{eff}}}
\newcommand\Bxif{\bar{\xi}}

\begin{document}

\title{Damping by branching: a bioinspiration from trees}

\author{Benoit~Theckes}
\email{theckes@ladhyx.polytechnique.fr}
\affiliation{Department of Mechanics, LadHyX, \'Ecole Polytechnique-CNRS, 91128 Palaiseau, France}
\author{Emmanuel~de~Langre}
\email{delangre@ladhyx.polytechnique.fr}
\affiliation{Department of Mechanics, LadHyX, \'Ecole Polytechnique-CNRS, 91128 Palaiseau, France}
\author{Xavier~Boutillon}
\email{boutillon@lms.polytechnique.fr}
\affiliation{Department of Mechanics, LMS, \'Ecole Polytechnique-CNRS, 91128 Palaiseau, France}
\date{\today}

\begin{abstract} Man-made slender structures are known to be sensitive to high levels of vibration, due to their flexibility, which often cause irreversible damage. In nature, trees repeatedly endure large amplitudes of motion, mostly caused by strong climatic events, yet with minor or no damage in most cases. A new damping mechanism inspired by the architecture of trees is here identified and characterized in the simplest tree-like structure, a Y-shape branched structure.
Through analytical and numerical analyses of a simple two-degree-of-freedom model, branching is shown to be the key ingredient in this protective mechanism that we call damping-by-branching. It originates in the geometrical nonlinearities so that it is specifically efficient to damp out large amplitudes of motion.
A more realistic model, using flexible beam approximation, shows that the mechanism is robust.
Finally, two bioinspired architectures are analyzed, showing significant levels of damping achieved via branching with typically $30\%$ of the energy being dissipated in one oscillation.
This concept of damping-by-branching is of simple practical use in the design of slender flexible structures.
\end{abstract}

\maketitle

\section{Introduction}

Vibrations in man-made structures are a central problem in mechanical engineering~\cite{DenHartog2007}.
They may result from external excitations such as wind, impacts, or earthquakes; or from internal excitations, such as a flow or moving parts.
Their consequences are numerous in terms of functionality losses due to wear, fatigue, or noise, to cite a few.
We distinguish here between low and high levels of vibration.
The former, where displacements are small in comparison with the characteristic size of the structure, may induce some of the long-term above-cited consequences.
The latter generally cause short-term failures and irreversible damage to the structure by fracture or plastic deformation~\cite{Collins1993}.
These large amplitudes of vibrations may be particularly expected in slender structures, or assemblages of them, due to their high flexibility.

In the most general framework of vibration analysis, the amplitude of motion results, on one hand, from the characteristics of the loading, and on the other hand, from the characteristics of the structure in terms of inertia, stiffness, and damping~\cite{Humar2002}.
Damping here refers to the capability of the structure to dissipate mechanical energy, whatever the physical mechanism involved (viscoelasticity, friction or interaction with a fluid).
A high level of damping in a structure is a standard way to avoid large-amplitude motions. 
This is generally achieved with passive techniques, such as the classical addition of tuned mass-damper systems~\cite[p. 119]{DenHartog2007}, or with active or semi-active means such as piezoelectric materials, magnetorheological fluids, shape memory alloys, or even simple hydraulic actuators in feedback or feedforward systems~\cite{Preumont2002}.
All these approaches have limits in terms of their range of acceptable deformations, displacements, or simply in terms of cost or maintenance.
But, more fundamentally, they stem from an ad-hoc corrective point of view, rather than from a consistent design perspective.
New approaches are needed, particularly for very slender and light structures, such as antennas, which may encounter large flow-induced amplitudes of vibration~\cite{Paidoussis2011}.

Nature may give insights into highly efficient mechanical solutions in vibration problems, for instance in shock-absorbing devices~\cite{Yoon2011}.
Interestingly, slender structures are ubiquitous in nature, particularly in plants.
Most of these plants are regularly submitted to natural flow excitations by wind or current causing vibrations~\cite{Langre2008}.
Although these vibrations contribute to some biological functions such as in seed or pollen dispersion, extreme events such as storms may cause dangerous large amplitudes of motion~\cite{Niklas1992}.
Therefore, in areas where intense flows are common, plants are likely to possess efficient damping strategies.

From a biomimetic point of view, the dynamical behaviour of trees, which has been extensively studied, is certainly a possible source of inspiration.
\begin{figure*}
	\centering
			\psfrag{pg}[br][Br]{$\phib$}
			\psfrag{pd}[bl][Bl]{$\phib$}
			\psfrag{m2g}[tr][tr][0.8]{$m_2$}
	 		\psfrag{m2d}[tl][tl][0.8]{$m_2$}
	 		\psfrag{m1}[Bc][Bc][0.8]{$m_1$}
			\psfrag{k2g}[tr][tr][0.8]{$k_2$}
	 		\psfrag{k2d}[tl][tl][0.8]{$k_2$}
	 		\psfrag{k1}[br][br][0.8]{$k_1$}
			\psfrag{l2g}[tr][tr][0.8]{$l_2$}
	 		\psfrag{l2d}[tl][tl][0.8]{$l_2$}
	 		\psfrag{l1}[br][bc][0.8]{$l_1$}
	 		\psfrag{b1}[cc][cc]{$\phi$}
	 		\psfrag{b2}[cl][cl]{$\phi$}
	 		\psfrag{a}[bc][bc]{$\theta$}
	\subfigure[{}]{ \includegraphics{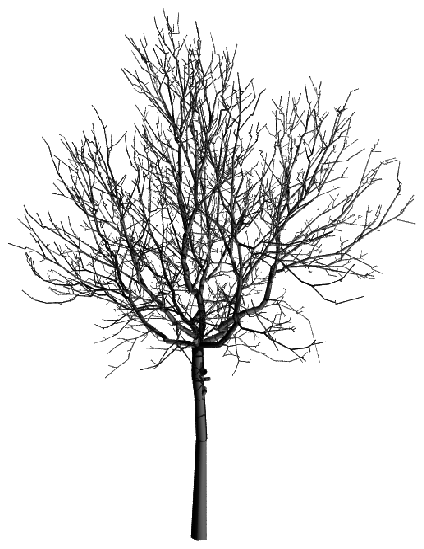} \label{fig:1a} }
	\subfigure[{}]{ \includegraphics{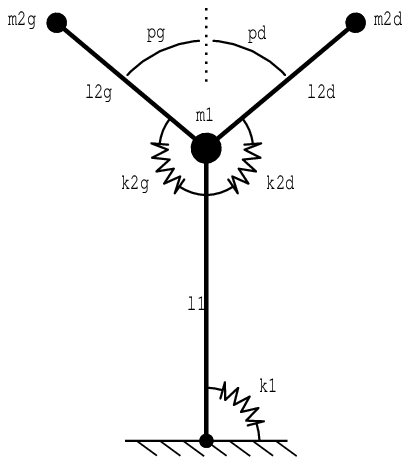} \label{fig:1b} } 
	\subfigure[{}]{ \includegraphics{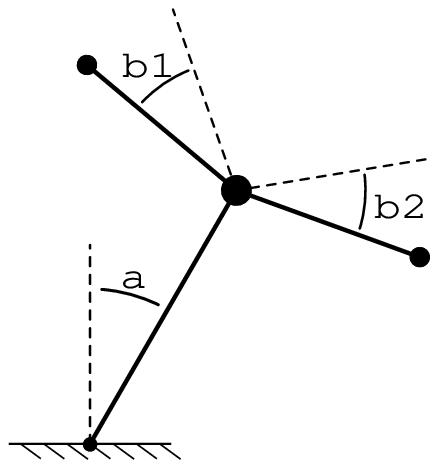} \label{fig:1c} }
	\caption{Branched geometries.
	(a) The walnut tree architecture analyzed by~\protect\citeasnoun{Rodriguez2008}.
	(b) and (c) Our Y-shaped spring-mass model of an elementary branched tree-like structure.
	\label{fig:1}}
\end{figure*}
\citeasnoun{Scannell1984} hinted that trees might possess a ``qualitative mechanical design principle [...] beneficial to the tree's survivability in conditions of strong atmospheric turbulence''.
\citeasnoun[p. 183]{Niklas1992} noted that ``experiments indicate that branching [...] dampens natural frequencies of vibration''.
At this point it is necessary to clarify what is generally agreed to cause damping in trees.
Firstly, the constitutive material, wood, is known to have a inherent viscoelastic behaviour causing dissipation, this itself has been the source of bioinspired material~\cite{Spatz2004}.
Secondly the aeroelastic interaction with the surrounding air causes forces in the opposite direction to the local velocity in the tree, thereby causing a strongly amplitude-dependent dissipation~\cite{blevins1990}.
Finally, when considering the overall motion of the tree by bending of the trunk, another mechanism is often described as ``structural damping''~\cite{Bruchert2003,Speck2004,James2006,Moore2008}.
This third mechanism refers to the possible transfer of mechanical energy from the trunk to the branches, where it will be eventually  dissipated by the two aforementioned aeroelastic and viscoelastic damping mechanisms~\cite{Sellier2009}.
\citeasnoun{Spatz2007} have suggested that the frequencies tuning of the branches plays a key role in this hypothesised energy transfer mechanism in trees.
In fact,~\citeasnoun{Rodriguez2008} analyzed the architecture of a walnut tree, figure~\ref{fig:1a}, and have shown that the modal frequencies are close and that the modal shapes are strongly localized in the architecture.
The former characteristic is classically favorable to modal energy exchanges in dynamical structures and the latter is consistent with localized energy transfers from the trunk to the branches.
In order to develop strategies for bioinspired designs of slender structures including an efficient damping effect for large amplitudes, it is crucial to clarify the mechanism involved in the energy transfer that many authors invoke.
Note that we do not intend to explore here the idea of the classical tuned-mass damper, which is based on a purely linear mechanism which, by definition, is not amplitude-dependent.
The aim of the present paper is therefore to identify and characterize the elementary mechanism causing nonlinear modal energy transfer and damping in a branched structure.

For this purpose, we first consider the simplest model of a branched dynamical system in Section~\ref{sec:Model}, a spring-mass model of a Y-shape.
Section~\ref{sec:FEM} shows, using a beam-finite-element model, that the main results of the previous section are also valid for a more realistic continuous structure of a Y-shape.
Based on these results, two illustrative designs of bioinspired slender structures exhibiting efficient damping-by-branching are proposed in Section~\ref{sec:bio}.
The generality and possible extensions of our approach are discussed in Section~\ref{sec:discuss}.

\section{Lumped-parameter model of a Y-shape} \label{sec:Model}

In order to reduce the dynamics of a branched structure to its simplest possible features, we treat the case of a spring-mass model of a Y-shape consisting of a trunk and two branches.
Since we are interested in the branching effect, viscous damping is introduced in the branches only.
The equations of motion are written with dimensionless variables and the dynamics is studied with an emphasis on the damping of the whole structure.

\subsection{Model}

The model consists of three massless rigid bars linked by rotational springs and supporting three masses, figure~\ref{fig:1b}.
The first bar, mimicking a trunk of length $l_1$, is linked to the ground by a rotational spring $k_1$ and supports a mass $m_1$.
The branches are two symmetrical bars of length $l_2$, each forming an angle $\phib$ with respect to the trunk axis.
Each branch is linked to the tip of the trunk by a rotational spring $k_2$, and supports a mass $m_2$.
The motion of the trunk is defined by the angle $\theta$, and we consider only the symmetrical motion of the branches defined by the angle $\phi$, figure~\ref{fig:1c}.
Note that this restriction is made in order to simplify the following dynamical analysis but has a small impact on the damping mechanism described in this section, as it will be seen in the following sections for much more complex models which have no such restrictions.
The kinetic energy is the sum of the kinetic energy of each mass,
\begin{multline}
	T = \frac{1}{2} \left[ \left( m_1 {l_1}^2 + 2 m_2 ( {l_1}^2 + 2 l_1 l_2 \cos(\phib + \phi) + {l_2}^2 ) \right) \dot{\theta}^2 \right.\\
	     \left. +~2 m_2 {l_2}^2 \dot{\phi}^2 \right].
\end{multline}
The potential energy is the sum of the potential energy of each spring,
\begin{equation}
	V = \dfrac{1}{2} \left( k_1 {\theta}^2 + 2 k_2 {\phi}^2 \right).
\end{equation}
The equations of motion are derived using $T$ and $V$ in the classical framework of Lagrangian dynamics~\cite{Humar2002}. They read
\begin{equation}
\begin{aligned}
J_{\theta} \ddot{\theta} + k_1 \theta &= 4 m_2 l_1 l_2 \left[ \dot{\theta} \dot{\phi} \sin(\phib + \phi) - \ddot{\theta} J_{\theta}^* \right] \\
2 m_2 {l_2}^2 \ddot{\phi}  + 2 k_2 \phi &= - 2 m_2 l_1 l_2 \dot{\theta}^2 \sin(\phib + \phi),
\end{aligned}
\label{eq:eqdim}
\end{equation}
where
\begin{equation}
	J_{\theta} = m_1 {l_1}^2 + 2 m_2 ( {l_1}^2 + 2 l_1 l_2 \cos{\phib} + {l_2}^2 )
\end{equation}
and $J_{\theta}^*(\phi) = \cos(\phib + \phi) - \cos{\phib}$.
The left-hand side of this system of equations represents two simple linear harmonic oscillators. Denoting the generalised displacement vector $\left[ \theta,\phi \right]$, the two corresponding normal modes of the system are directly $\left[ 1,0 \right]$ and $\left[ 0,1 \right]$ since there is no linear coupling between $\theta$ and $\phi$. The two modal angular frequencies are respectively
\begin{equation}
	  {\omega_1}^2 = \frac{k_1}{J_{\theta}}, \text{~and~~~} {\omega_2}^2 = \frac{2 k_2}{2 m_2 {l_2}^2}.
	  \label{eq:freq}
\end{equation}
The first mode consists of motion involving $\theta$ only, and the second mode involving $\phi$ only.
Therefore, in the following, they are referred to as the trunk mode and the branch mode respectively.
These two modes are coupled by the nonlinear terms of the right-hand side of \ref{eq:eqdim}, representing the geometric nonlinearities.

A dimensional analysis reveals the existence of four dimensionless parameters describing the dynamics of the model.
We chose the dimensionless time $\tau = \omega_1 t$, the branching angle $\phib$, the ratio of angular frequencies $\Omega = {\omega_2}/{\omega_1}$, and $\Gamma$, the ratio between the inertial terms of the branch mode and the trunk mode, multiplied by the length ratio ${l_1}/{l_2}$,
\begin{equation}
		\Gamma = \frac{2 m_2 {l_2}^2}{J_{\theta}} \frac{l_1}{l_2} = \frac{2 m_2 l_1 l_2}{J_{\theta}}.
 		\label{eq:gama}
\end{equation}
The dynamics is described by the variables $\Theta(\tau) = \theta(t) \sqrt{{l_1}/{l_2}}$ and $\Phi(\tau) = \phi(t)$.
As mentioned earlier, we introduce energy dissipation in the form of a viscous damping rate $\xib$ in the branch mode only.
The dimensionless equations of motion are
\begin{equation}
\begin{aligned}
		\ddot{\Theta} + \Theta &= 2 \Gamma \left[ \dot{\Theta} \dot{\Phi} \sin(\phib + \Phi) - \ddot{\Theta} J_{\theta}^* \right], \\
			        \ddot{\Phi} + 2\Omega\xib\dot{\Phi} + \Omega^2\Phi &= -\dot{\Theta}^2 \sin(\phib + \Phi).
\end{aligned}
		\label{eq:eqadim}
\end{equation}
Note that $J_{\theta}^*(\phi) = J_{\theta}^*(\Phi)$ since $\Phi(\tau) = \phi(t)$.
The dimensionless total mechanical energy is
\begin{multline}
	E(\tau) = \frac{1}{2} \left[ \left( 2 \Gamma \left( \cos(\phib+\Phi) - \cos{\phib} \right) + 1 \right) {\dot \Theta }^2 + \Theta^2 \right. \\
	           \left. +~\Gamma \left( {\dot \Phi}^2 + {\Omega}^2 \Phi^2 \right) \right]. \label{eq:energy}
\end{multline}
Since the two modes are coupled by nonlinear terms, energy can be exchanged between them. In this case, the dissipation in the branch mode may damp the energy received from the trunk mode, resulting in an effective damping of the whole structure.

\subsection{Damping criterion} \label{sect:crit}

In the following, we examine the free vibrations following an initial condition
\begin{equation}
	  \left[ \Theta(0), \dot{\Theta}(0), \Phi(0), \dot{\Phi}(0) \right] = \left[\Theta_0, 0 , 0, 0\right],
 		\label{eq:CI}
\end{equation}
such that the energy is located in the undamped trunk mode only.
This will allow us to easily demonstrate damping by nonlinear modal energy transfer, if any, in a purely linear framework, energy would remain in the undamped trunk mode with no way of being dissipated.
The amplitude of the initial condition, $\Theta_0$, determines the initial energy $E(0) = E_0$, using \ref{eq:energy}.
For the sake of clarity, the energy $E$ is normalized so that the initial energy $E_0$ is $1$ when $\Theta_0 = \pi/2$ corresponding to a horizontal trunk initial condition. Note that ground interaction is here neglected.

During free oscillations, a part of the energy transferred from the trunk mode to the branch mode is dissipated. The total energy decay over the first period of the trunk mode is $\Delta E = E_0 - E(2\pi)$ so that the effective damping rate of the whole structure can be defined as
\begin{equation}
				\xif = \frac{1}{4\pi} \frac{\Delta E}{E_0}.
  		\label{eq:zeff}
\end{equation}
The effective damping rate, $\xif$, is commonly related to the quality factor by $Q_{\rm{eff}} = 1 / {(2 \xif)}$.
Note that $\xif$ represents the dissipation of the whole structure and not that of the trunk mode. In fact, studying exclusively the trunk mode damping is not appropriate since energy transfer can be reciprocal from the branch mode to the trunk mode as well, as will be seen in~figure~\ref{fig:Etot}.
The total energy decay $\Delta E$ is given by the work of the damping term of the branch mode equation over one period of the trunk mode:
\begin{equation}
				\Delta E = \frac{8}{\pi^2} \Gamma \int_0^{2\pi} 2 \Omega \xib  \dot{\Phi}^2 \rm{d}\tau .
		\label{eq:DE}
\end{equation}
Here, the coefficient $8/{\pi^2}$ comes from the normalization chosen for $E$.
We analyze now the effect of the initial energy $E_0$ and the design parameters $\phib$, $\xib$, $\Omega$, and $\Gamma$ on the effective damping, $\xif$.

\subsection{Energy transfer by internal resonance} \label{sect:analy}

In this section, we consider a low initial energy level so that $\Theta_0= \varepsilon$, where $\varepsilon \ll 1$ is a small parameter.
The harmonic balance method~\cite{Nayfeh1979} is used with the angles $\Theta$ and $\Phi$ developed as power series of $\varepsilon$,
\begin{eqnarray}
	  \Theta(\tau) &=& \varepsilon \Theta_1(\tau) + \varepsilon^2 \Theta_2(\tau) + \cdots, \label{eq:teta} \\
		\Phi(\tau) &=& \varepsilon \Phi_1(\tau) + \varepsilon^2 \Phi_2(\tau) + \cdots . \label{eq:phi}
\end{eqnarray}
The initial condition \ref{eq:CI} requires that
\begin{equation}
	  \Theta_1(0) = 1, \text{~and~~~} \Theta_2(0) = \Phi_1(0) = \Phi_2(0) = 0. 
 		\label{eq:CI2}
\end{equation}
Substituting \ref{eq:teta} and \ref{eq:phi} in the dynamical equations \ref{eq:eqadim}, and using \ref{eq:CI2}, the first-order terms are
\begin{equation}
	  \Theta_1 = \cos\tau, \text{~and~~~} \Phi_1 = 0.
 		\label{eq:espi1}
\end{equation}
The second-order terms satisfy respectively
\begin{equation}
		\Theta_2 = 0, \text{~and~~} \ddot{\Phi}_2 + 2\Omega\xib\dot{\Phi}_2 + \Omega^2\Phi_2 = -\dot{\Theta_1}^2 \sin\phib. \label{eq:epsi2}
\end{equation}
Therefore, for small angles, \ref{eq:eqadim} reduces to
\begin{equation}
   \ddot{\Phi} + 2\Omega\xib\dot{\Phi} + \Omega^2\Phi = - \Theta_0^2 \sin\phib \frac{(1 - \cos2\tau)}{2}. 
	 \label{eq:reso}
\end{equation}
This is the equation of a simple harmonic damped oscillator, driven by an harmonic force, that can be analytically solved~\cite{Humar2002}.
A resonance exists at $\Omega = 2$; since $\Omega$ is the frequency ratio of the two modes, this is classically referred to as a $1$:$2$ internal resonance~\cite{Nayfeh1979}.
In the following, we will discuss the influence of $\Omega$ near this particular value.
A general result for a forced damped oscillator is that the amplitude of motion is proportional to the amplitude of the driving force.
As can be seen in \ref{eq:reso}, the amplitude of the driving force is proportional to $\Theta_0^2 \sin\phib$, and therefore to $E_0 \sin\phib$.
The effective damping $\xif$, defined by \ref{eq:zeff} and \ref{eq:DE}, can therefore be simply expressed as
\begin{equation}
		\xif = E_0 \Gamma \sin^2\phib ~\Bxif(\xib,\Omega).
		\label{eq:ana}
\end{equation}
Remarkably, \ref{eq:ana} shows that $\xif$ increases linearly with the initial energy $E_0$: such a nonlinear damping proportional to the energy is typical of an oscillator following the generic equation: $\ddot{\Theta} + \kappa \dot{\Theta}^3 + \Theta = 0$~\cite{Nayfeh1979}.
Besides, \ref{eq:ana} shows that $\xif$ is proportional to $\sin^2\phib$ so that the effective damping is maximal for a branching angle $\phib = \pi / 2$, corresponding to a T-shaped structure.
Conversely, for a non-branched structure, where $\phib = 0$ or $\pi$, the effective damping is zero.
The effective damping is also proportional to the relative modal mass ratio $\Gamma$.
\begin{figure}[h!]
	\centering
	  \psfrag{x}[b][b]{$\Omega$}
  	\psfrag{y}[l][l]{$\xib$}
  	\psfrag{z}[cc][cc][1.2]{$\Bxif$}	
				\psfrag{y0}[br][r][0.8]{$0$}
				\psfrag{y1}[r][r][0.8]{$0.5$}
				\psfrag{y2}[r][r][0.8]{$1$}
					\psfrag{x0}[l][][0.8]{$1$}
					\psfrag{x1}[][][0.8]{$2$}
					\psfrag{x2}[][][0.8]{$3$}
				\psfrag{yb0}[bl][cl][0.8]{$0$}
				\psfrag{yb1}[cl][cl][0.8]{$0.05$}
				\psfrag{yb2}[cl][cl][0.8]{$0.1$}
				\psfrag{yb3}[cl][cl][0.8]{$0.15$}
				\psfrag{yb4}[cl][cl][0.8]{$0.2$}
  \includegraphics{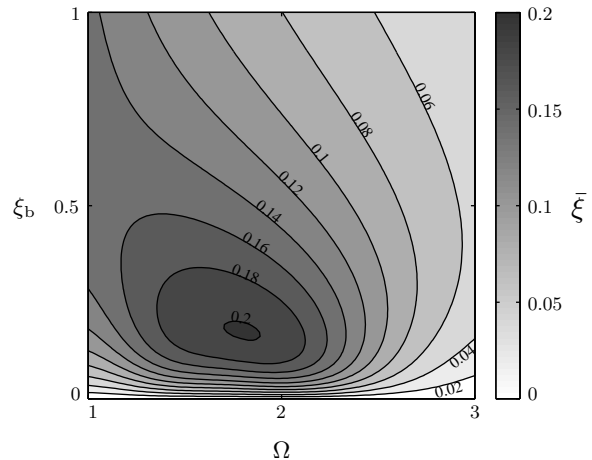}  
  \caption{Normalized effective damping $\Bxif$, \ref{eq:ana}, brought by branching,
  as a function of the branch mode damping $\xib$, and the branch/trunk modal frequency ratio $\Omega$.
  A high damping is found near the $1$:$2$ internal resonance, i.e. at $\Omega = 2$.
	\label{fig:Xibar}}
\end{figure}
The dependency of $\xif$ on $\Omega$ and $\xib$ is shown in figure~\ref{fig:Xibar} as a contour-map of the normalised effective damping $\Bxif$, computed using a mathematic symbolic software to solve \ref{eq:reso} for $\Phi$, and then using successively \ref{eq:DE}, \ref{eq:zeff}, and \ref{eq:ana} for $\Bxif$.
We observe that a significant level of damping is present over a wide range of parameter values.
The effective damping shows a maximum for branch damping near $0.2$ and frequency ratio near $2$.
Accordingly, the values $\phib = \pi / 2$, $\xib = 0.2$, $\Omega = 2$ and $\Gamma = 0.2$ will be used as a reference in the remainder of this paper.

As expected, we observe in figure~\ref{fig:Xibar} that there is no effective damping for $\xib = 0$ since mechanical energy cannot be dissipated in the structure.
Interestingly, for any arbitrary small value of $\xib$, the effective damping is finite.
In a purely linear framework, the effective damping would be zero for any value of $\xib$, since the total energy would be confined to the trunk mode, without any possible transfer to the branch mode where dissipation occurs. 
In other words, the effective damping is due to the geometric nonlinearities.

\subsection{Effects of the design parameters} \label{sect:num}

In order to obtain the full dynamics and the corresponding effective damping at any energy level with an emphasis on the effects of the design parameters $\phib$, $\xib$, $\Omega$ and $\Gamma$, the dynamical system~\ref{eq:eqadim} is now solved numerically by means of an fourth-order explicit Runge-Kutta temporal scheme.
\begin{figure}[!ht]
	\centering
	\psfrag{x}[b][b]{$\tau / 2\pi$}
	\psfrag{y}[ct][ct]{$E$, $E_{\Theta}$, $E_{\Phi}$}
	\psfrag{yb}[ct][ct]{$\Theta$, $\Phi$}
		\psfrag{de}[cl][cl]{$\Delta E$}
			  \psfrag{x0}[l][][0.8]{$0$}
				\psfrag{x1}[][][0.8]{$1$}
				\psfrag{x2}[][][0.8]{$2$}
				\psfrag{x3}[][][0.8]{$3$}
				\psfrag{y0}[br][r][0.8]{$0$}
				\psfrag{y1}[r][r][0.8]{$0.5$}
				\psfrag{y2}[r][r][0.8]{$1$}
				\psfrag{y20}[br][br][0.8]{$-\pi/2$}
				\psfrag{y21}[r][r][0.8]{$0$}
				\psfrag{y22}[tr][tr][0.8]{$\pi/2$}
	\includegraphics{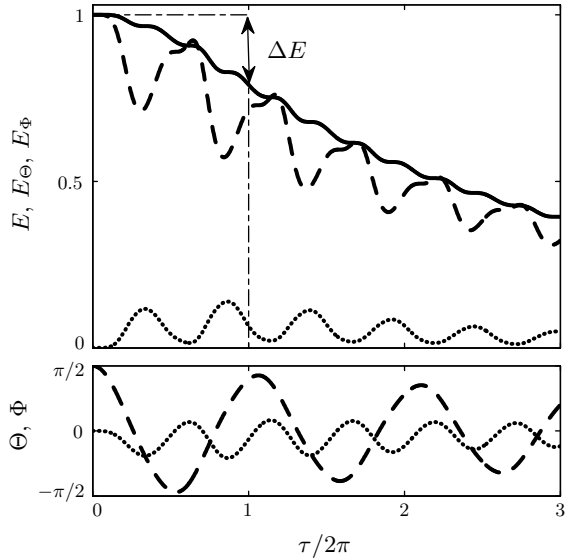}
	\caption{Typical evolution of the total energy, $E$ (---), and modal energies, $E_{\Theta}$ (-~-~-) and
	 $E_{\Phi}$ ($\cdots$) with the respective evolution of the trunk angle, $\Theta$ (-~-~-), and branch angle, $\Phi$ ($\cdots$), of the spring-mass model of a Y-shape, as a function of time over three periods of the trunk mode.
	 The initial energy is $E_0 = 1$, i.e $\Theta_0 = \pi/2$, in the trunk mode only.
	 The total energy decreases as a consequence of the energy nonlinearly transferred to the damped branch mode.
	 The design parameters are set to $\phib = \pi/2$, $\xib = 0.2$, $\Omega = 2$, and $\Gamma = 0.2$.
	\label{fig:Etot}}
\end{figure}
As a typical example, figure~\ref{fig:Etot} shows, for an initial energy level $E_0 = 1$, the evolution of the total energy $E$ and of the modal energies
\begin{eqnarray}	
  E_{\Theta} &=& \dfrac{4}{\pi^2} \left( {\dot \Theta }^2 + \Theta^2  \right), \\
	E_{\Phi} &=& \dfrac{4}{\pi^2} \Gamma \left( {\dot \Phi}^2 + {\Omega}^2 \Phi^2 \right). \label{eq:E1E2}
\end{eqnarray}
Note that the total energy $E$, \ref{eq:energy}, is the sum of $E_{\Theta}$, $E_{\Phi}$ and a nonlinear energy term.
The energy exchange between the two modes is clearly shown.
Since energy is dissipated in the branch mode, the total energy decays at an effective damping rate $\xif$.
Figure~\ref{fig:Zeff_vs_E0} shows the $E_0$-dependency of this effective damping $\xif$, in comparison with the analytical prediction of the previous section.
\begin{figure}[!ht]
	\centering
	\psfrag{x}[b][b]{$E_0$}
	\psfrag{y}[l][l]{$\xif$}
		\psfrag{a1}[bl][bl][0.8]{$.10^{-2}$}
				\psfrag{x0}[l][][0.8]{$0$}
				\psfrag{x1}[][][0.8]{$0.5$}
				\psfrag{x2}[][][0.8]{$1$}
				\psfrag{y0}[br][r][0.8]{$0$}
				\psfrag{y1}[r][r][0.8]{$1$}
				\psfrag{y2}[r][r][0.8]{$2$}
				\psfrag{y3}[r][r][0.8]{$3$}
				\psfrag{y4}[r][r][0.8]{$4$}
	\includegraphics[scale=0.8]{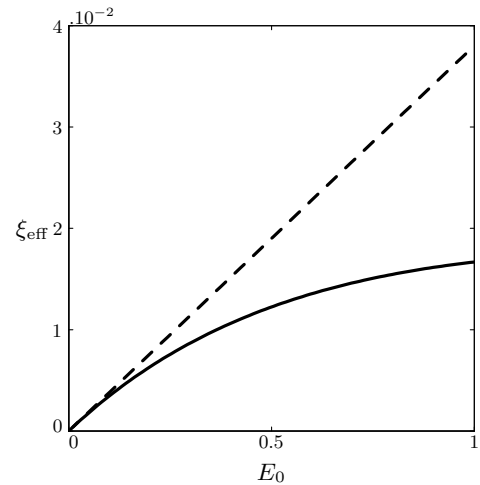}
	\caption{Effect of the initial energy level, $E_0$, on the effective damping, $\xif$, of the spring-mass model of a Y-shape:
	(-~-~-) analytical effective damping from the low energy approximation, \ref{eq:ana}; (---) numerical effective damping from the full dynamics integration. The design parameters are set to the same values as in figure~\ref{fig:Etot}.
  \label{fig:Zeff_vs_E0}}
\end{figure}
As expected, the analytical approach corresponds to the limit of the numerical solution as $E_0$ tends to zero.
As $E_0$ increases, the analytical approach increasingly overestimates the numerical effective damping.
However, the ratio $\xif / {E_0}$ ---constant in the analytical approach--- remains finite: this constitutes the essential effect of branching on damping.

\begin{figure}[!ht]
	\centering
		\psfrag{y}[][l]{$\dfrac{\xif}{E_0}$}
		\psfrag{a1}[bl][bl][0.7]{$.10^{-2}$}
		\psfrag{x1}[c][b]{$\phib$}
				\psfrag{x10}[tl][t][0.7]{$0$}
				\psfrag{x11}[tc][t][0.7]{$\pi/2$}
				\psfrag{x12}[tc][t][0.7]{$\pi$}
		\psfrag{x2}[c][b]{$\xib$}
				\psfrag{x20}[tl][t][0.7]{$0$}
				\psfrag{x21}[tc][t][0.7]{$0.5$}
				\psfrag{x22}[tc][t][0.7]{$1$}
		\psfrag{x3}[c][b]{$\Omega$}
				\psfrag{x30}[tl][t][0.7]{$1$}
				\psfrag{x31}[tc][t][0.7]{$2$}
				\psfrag{x32}[tc][t][0.7]{$3$}
		\psfrag{x4}[c][b]{$\Gamma$}
				\psfrag{x40}[tl][t][0.7]{$0$}
				\psfrag{x41}[tc][t][0.7]{$0.2$}
				\psfrag{x42}[tc][t][0.7]{$0.4$}
					\psfrag{y40}[b][r][0.7]{$0$}
					\psfrag{y41}[][r][0.7]{$2$}
					\psfrag{y42}[][r][0.7]{$4$}
					\psfrag{y43}[][r][0.7]{$6$}			
					\psfrag{y44}[][r][0.7]{$8$}
  			\psfrag{y10}[b][r][0.7]{$0$}
				\psfrag{y11}[][r][0.7]{$1$}
				\psfrag{y12}[][r][0.7]{$2$}
				\psfrag{y13}[][r][0.7]{$3$}			
				\psfrag{y14}[][r][0.7]{$4$}
	\subfigure[{}]{ \includegraphics[scale = 0.6]{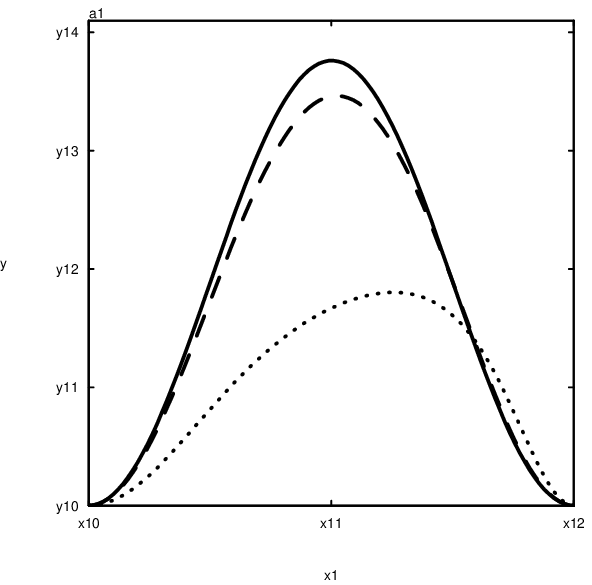}	\label{fig:Design_a} } 
	\subfigure[{}]{ \includegraphics[scale = 0.6]{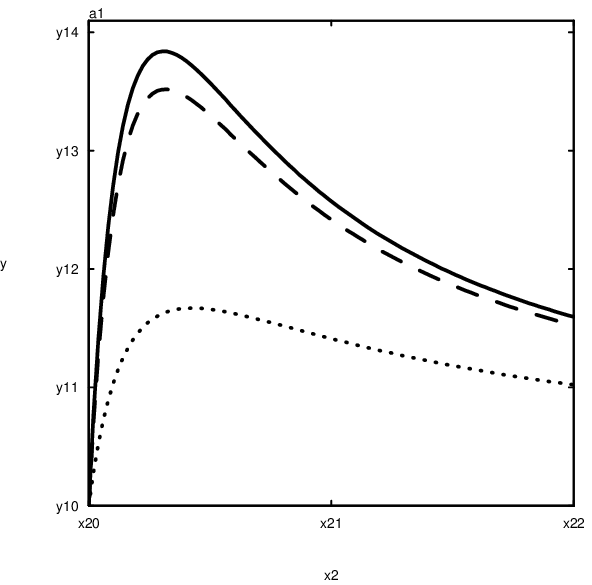}	\label{fig:Design_b} } \\
	\subfigure[{}]{ \includegraphics[scale = 0.6]{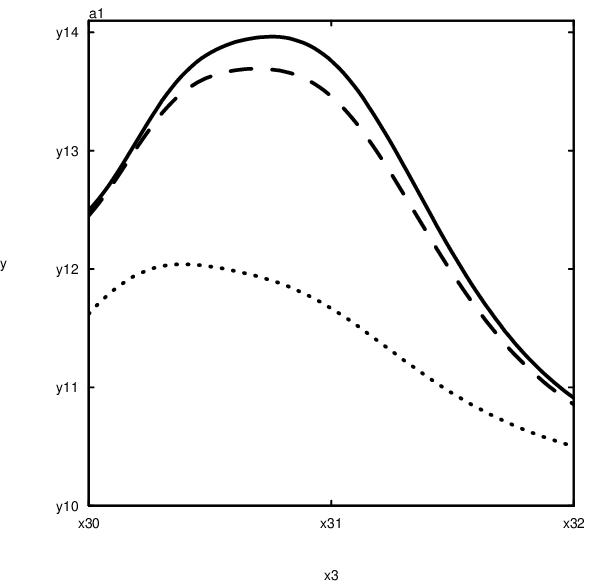} \label{fig:Design_c} } 
	\subfigure[{}]{ \includegraphics[scale = 0.6]{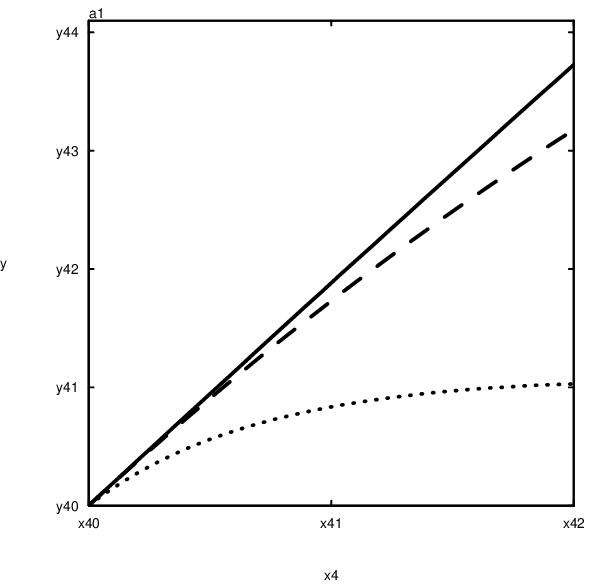} \label{fig:Design_d} }
	\caption{Effects of the design parameters on the effective damping scaled by the initial energy, $\xif / E_0$, 
	of the spring-mass model of a Y-shape: $E_0 = 0.01$ (---); $E_0 = 0.1$ (-~-~-); and $E_0 = 1$ ($\cdots$).
	Unless varied, the design parameters values are: $\phib = \pi/2$, $\xib = 0.2$, $\Omega = 2$, and $\Gamma = 0.2$.
  (a) Effect of the branching angle $\phib$.
  (b) Effect of the branch mode damping $\xib$.
  (c) Effect of the branch/trunk modal frequency ratio $\Omega$.
  (d) Effect of the branch/trunk modal mass ratio $\Gamma$.
  \label{fig:Design}}
\end{figure}

The influences of the design parameters $\phib$, $\xib$, $\Omega$, and $\Gamma$ on the effective damping, scaled by the initial energy, $\xif / {E_0}$, are represented in figures \ref{fig:Design}(a)--(d) for three initial energy levels.
As expected, the analytical and numerical approaches yield identical results for low energy levels and therefore are represented by the same curve for $E_0 = 0.01$.
Consistently with figure~\ref{fig:Zeff_vs_E0}, we observe that the analytical approach overestimates the numerical effective damping as $E_0$ increases.
In figure~\ref{fig:Design_a}, at low energy levels, the effective damping is proportional to $\sin^2\phib$ as predicted by the analytical approach \ref{eq:ana}.
The optimal branching angle $\phib$ shifts from $\pi/2$ to slightly higher values when the initial energy level increases.
Therefore, in order to obtain an optimum effective damping at any energy, a good compromise would be to set the branching angle $\phib$ between $\pi/2$ and $2\pi/3$.

Figure~\ref{fig:Design_b} shows that a significant effective damping is created by a large range of branch mode damping, $\xib$, as was also found in the low-energy analytical approach.
Analogously, the typical shape of the $\Omega$-dependency at low energy is also conserved when the energy increases, figure~\ref{fig:Design_c}.
The simple $\Gamma$-dependency on the effective damping is shown in figure~\ref{fig:Design_d}: the modal mass ratio $\Gamma$ has to be maximal in order to get the highest possible effective damping at any energy level.

\section{Finite-element model of a Y-shape} \label{sec:FEM}

The damping-by-branching mechanism described in the preceding section is now analysed in the case of a more realistic continuous beam structure of a Y-shape.
The same approach is used to demonstrate the effective damping: initial energy in the trunk, dissipation in the branches and effective damping evaluated by the total energy loss over one period of the trunk mode.
Note that this model incorporates several differences with the previous one: a very large number of modes, symmetric and non-symmetric modes, non-localized mass and stiffness.

\subsection{Model}

\begin{figure}[!ht]
	\centering
			\psfrag{a1}[bl][]{$\phib$}
			\psfrag{a2}[br][]{$\phib$}
			\psfrag{d3}[bc][b][0.8]{$\lambda$}
	 		\psfrag{d2}[bc][b][0.8]{F}
	~~~~~~ \subfigure[{}]{ \includegraphics{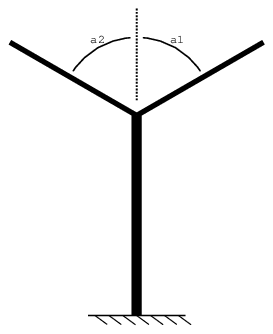}	\label{fig:Y_FEM_a} } ~
	\subfigure[{}]{ \includegraphics{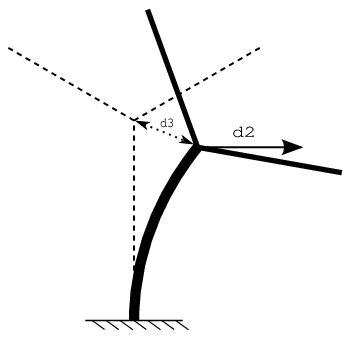}	\label{fig:Y_FEM_b} } \\
	\subfigure[{}]{ \includegraphics{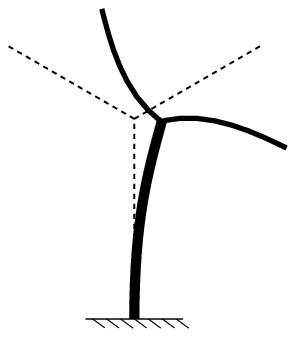}	\label{fig:Y_FEM_mode1} }
	\subfigure[{}]{ \includegraphics{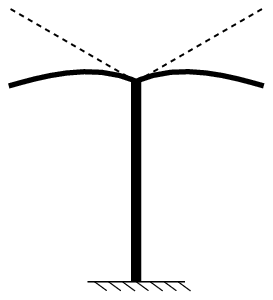}	\label{fig:Y_FEM_mode2} }
	\caption{The continuous model of a Y-shape.
	(a) Geometry.
	(b) Static initial condition.
	(c) Trunk mode.
	(d) Damped branch mode.
	\label{fig:Y_FEM}}
\end{figure}

The model consists of three assembled beams, figure~\ref{fig:Y_FEM_a}. 
Each beam has a uniform circular cross-section and is made of a linearly elastic, isotropic and homogeneous material.
The trunk, of length $l_1$ and diameter $d_1$, is clamped at the base. 
Two symmetrical branches, each of length $l_2$ and diameter $d_2$, are clamped at the tip of the trunk so that they each form an angle $\phib$ with the trunk direction.
As in Section~\ref{sec:Model}, we analyze the free vibrations of the structure. 
By analogy with \ref{eq:CI}, the initial condition is an initial deformation resulting from a horizontal static pull on the tip of the trunk, figure~\ref{fig:Y_FEM_b}.
This initial energy $E_0$ is normalized so that it is equal to $1$ when the deflection of the trunk is equal to its length such that $\lambda = l_1$.
To solve the equations of motion, finite-element computations are performed using the CASTEM v.3M software \cite{verpeaux1988castem}.
In order to take into account large displacements, a step-by-step procedure is needed implying an update of the deformed configuration.
Timoshenko-beam elements are used taking into account the rotational inertia of the sections of the beams.
Ten mesh-elements per beam were chosen according to a convergence test on the initial energy repartition on modes.
It appears that this number is sufficient to describe the full dynamics of the system.

The first two modal shapes are given by modal analysis and are shown in figures \ref{fig:Y_FEM_mode1} and \ref{fig:Y_FEM_mode2}.
We observe that the branch mode, figure~\ref{fig:Y_FEM_mode2}, involves only branches bending while the trunk mode, figure~\ref{fig:Y_FEM_mode1}, involves both trunk and branch deformations.
This modal localization and the initial condition implies that the branch mode has no initial energy.
Note that the initial energy is mainly in the trunk mode but can be shared with other modes than the branch mode.
By analogy with Section~\ref{sec:Model}, two dimensionless parameters are chosen: the frequency ratio $\Omega = \omega_2/\omega_1$ and a mass ratio $\Gamma = (l_1 m_2) / (l_2 m_1)$, where $\omega_1$, $\omega_2$, and $m_1$, $m_2$ are the modal angular frequencies and the modal masses of the trunk mode and branch mode respectively.

As in Section~\ref{sec:Model}, energy dissipation is introduced artificially on the branch mode only by a damping matrix derived from the mass matrix of the finite-element model with a damping rate denoted $\xib$.
The resulting effective damping mechanism is studied with the same definition of the effective damping rate, $\xif$, as in Section~\ref{sect:crit}, and for the same reference values of the design parameters $\phib = \pi/2$, $\xib = 0.2$, $\Omega = 2$, and $\Gamma = 0.2$.

\subsection{Results}

\begin{figure}[!ht]
	\centering
	\psfrag{x}[b][b]{$E_0$}
	\psfrag{y}[l][l]{$\xif$}
		\psfrag{a1}[bl][bl][0.8]{$.10^{-2}$}
				\psfrag{x0}[l][][0.8]{$0$}
				\psfrag{x1}[][][0.8]{$0.5$}
				\psfrag{x2}[][][0.8]{$1$}
				\psfrag{y0}[br][r][0.8]{$0$}
				\psfrag{y1}[r][r][0.8]{$1$}
				\psfrag{y2}[r][r][0.8]{$2$}
				\psfrag{y3}[r][r][0.8]{$3$}
				\psfrag{y4}[r][r][0.8]{$4$}
	\includegraphics[scale=0.8]{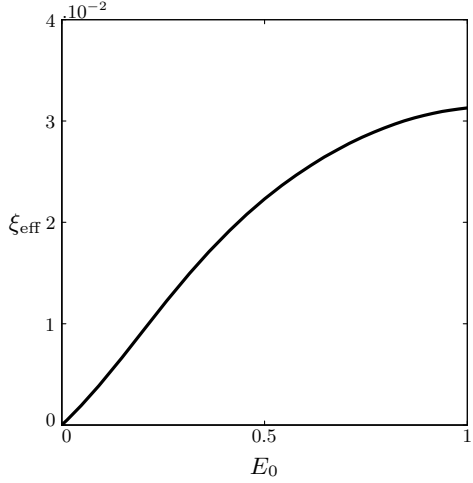}
	\caption{Effect of the initial energy level, $E_0$, on the effective damping, $\xif$, of the continuous model of a Y-shape.
	The design parameters are set to $\phib = \pi/2$, $\xib = 0.2$, $\Omega = 2$, and $\Gamma = 0.2$.
  \label{fig:FEM_Zeff_vs_E0}}
\end{figure}

The simulated dynamics of the continuous Y-shape yields a similar time evolution of the total energy to that of the lumped parameter model, figure~\ref{fig:Etot}.
The corresponding effective damping rate is plotted in figure~\ref{fig:FEM_Zeff_vs_E0} as a function of the normalized initial energy $E_0$.
As in figure~\ref{fig:Zeff_vs_E0}, the effective damping rate increases with the initial energy level, quasi-linearly at first and reaching several percent for high levels of initial energy. 
This is a first indication of the robustness of the effect of branching on damping in a more realistic structure.
\begin{figure}[!ht]
	\centering
		\psfrag{y}[][l]{$\dfrac{\xif}{E_0}$}
		\psfrag{a1}[bl][bl][0.7]{$.10^{-2}$}
		\psfrag{x1}[c][b]{$\phib$}
				\psfrag{x10}[tl][t][0.7]{$0$}
				\psfrag{x11}[t][t][0.7]{$\pi/2$}
				\psfrag{x12}[t][t][0.7]{$\pi$}
		\psfrag{x2}[c][b]{$\xib$}
				\psfrag{x20}[tl][t][0.7]{$0$}
				\psfrag{x21}[tc][t][0.7]{$0.5$}
				\psfrag{x22}[tc][t][0.7]{$1$}
		\psfrag{x3}[c][b]{$\Omega$}
				\psfrag{x30}[tl][t][0.7]{$1$}
				\psfrag{x31}[tc][t][0.7]{$2$}
				\psfrag{x32}[tc][t][0.7]{$3$}
				  \psfrag{y33}[cr][cr][0.7]{$9$}	
		\psfrag{x4}[c][b]{$\Gamma$}
				\psfrag{x40}[tl][t][0.7]{$0$}
				\psfrag{x41}[tc][t][0.7]{$0.2$}
				\psfrag{x42}[tc][t][0.7]{$0.4$}
					\psfrag{y40}[br][r][0.7]{$0$}
					\psfrag{y41}[r][r][0.7]{$4$}
					\psfrag{y42}[r][r][0.7]{$8$}
					\psfrag{y43}[r][r][0.7]{$12$}
				  \psfrag{y10}[br][r][0.7]{$0$}
				  \psfrag{y11}[r][r][0.7]{$3$}
				  \psfrag{y12}[r][r][0.7]{$6$}	
	\subfigure[{}]{ \includegraphics[scale=0.6]{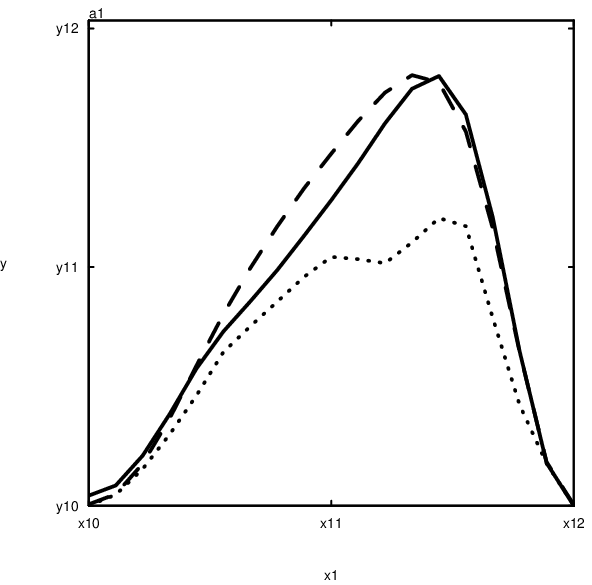}	\label{fig:FEM_Design_a} } 
	\subfigure[{}]{ \includegraphics[scale=0.6]{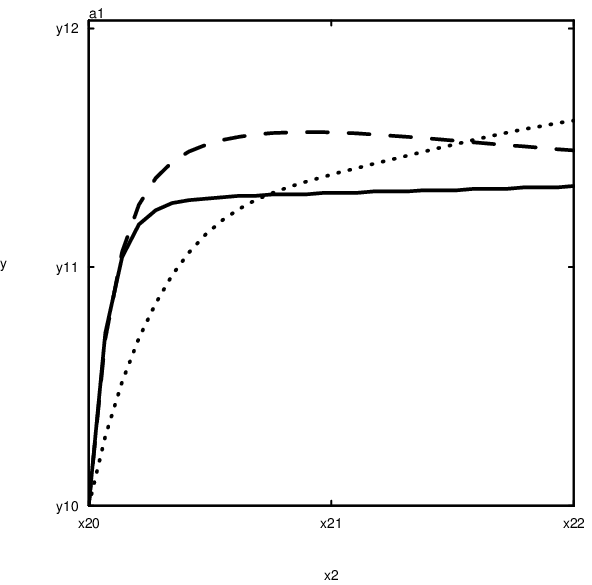}	\label{fig:FEM_Design_b} } \\
	\subfigure[{}]{ \includegraphics[scale=0.6]{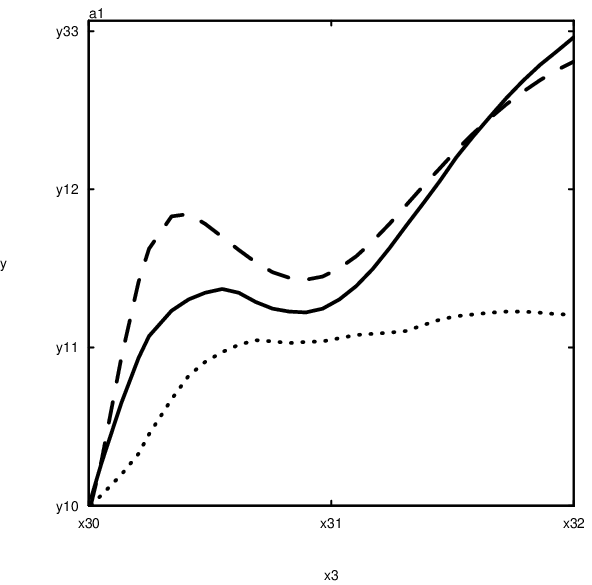} \label{fig:FEM_Design_c} } 
	\subfigure[{}]{ \includegraphics[scale=0.6]{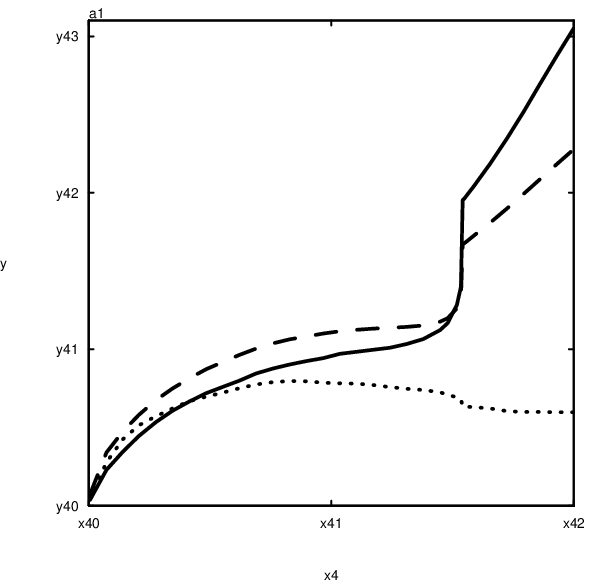} \label{fig:FEM_Design_d} }
	\caption{Effects of the design parameters on the effective damping scaled by the initial energy, $\xif / E_0$, 
	of of the continuous model of a Y-shape:: $E_0 = 0.01$ (---); $E_0 = 0.1$ (-~-~-); and $E_0 = 1$ ($\cdots$).
	Unless varied, the design parameters values are: $\phib = \pi/2$, $\xib = 0.2$, $\Omega = 2$, and $\Gamma = 0.2$.
  (a) Effect of the branching angle $\phib$.
  (b) Effect of the branch mode damping $\xib$.
  (c) Effect of the branch/trunk modal frequencies ratio $\Omega$.
  (d) Effect of the branch/trunk modal mass ratio $\Gamma$.
  \label{fig:FEM_Design}}
\end{figure}
As in Section~\ref{sec:Model}, the effects of the design parameters $\phib$, $\xib$, $\Omega$, and $\Gamma$ on the effective damping scaled by the initial energy, $\xif / {E_0}$, are represented in figures \ref{fig:FEM_Design}(a)--(d) for three initial energy levels.
Some differences appear for the continuous model as expected.
Firstly, the branching effect on the effective damping is clearly maximal for larger branching angles, $\phib \approx 2\pi/3$, rather than $\pi/2$, figure~\ref{fig:FEM_Design_a}.
Secondly, the effective damping stabilizes or even slightly increases with the branch mode damping $\xib$, figure~\ref{fig:FEM_Design_b}, instead of decreasing after $\xib \approx 0.2$.
Thirdly, the effective damping is higher at a modal frequency ratio $\Omega = 3$ than $\Omega = 2$, figure~\ref{fig:FEM_Design_c}, suggesting a richer pattern of internal resonances.
Singularly, the effective damping disappears for $\Omega = 1$, which is a striking difference with the classical tuned-mass damper model.
Finally, the effective damping increases with the modal mass ratio $\Gamma$ but in a more complex way, figure~\ref{fig:FEM_Design_d}.\\
In conclusion, it appears that the main features of the damping-by-branching mechanism are still present in this more realistic Y-shape structure.

\section{Two bioinspired branched structures} \label{sec:bio}
\begin{figure}[!ht]
	\centering
	\subfigure[{}]{ \includegraphics{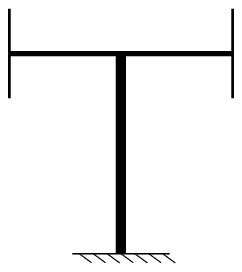}	\label{fig:T} } \\
	\subfigure[{}]{ \includegraphics{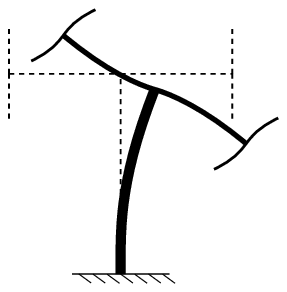}	\label{fig:T_1} }
	\subfigure[{}]{ \includegraphics{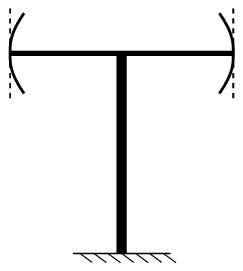}	\label{fig:T_2} } \\
	\subfigure[{}]{ \includegraphics{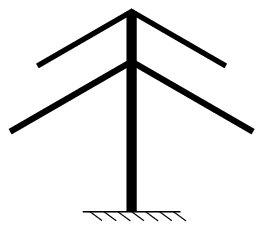}	\label{fig:S} } \\
	\subfigure[{}]{ \includegraphics{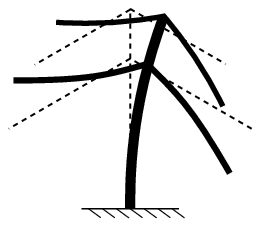}	\label{fig:S_1} }
	\subfigure[{}]{ \includegraphics{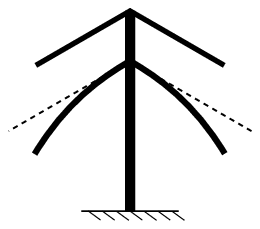}	\label{fig:S_2} }
	\subfigure[{}]{ \includegraphics{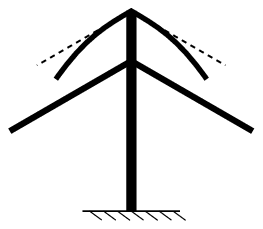}	\label{fig:S_3} }
	\caption{Two bioinspired branched structures.
	(a) A two-orders ramified T-shaped structure, and its modes of interest: (b) Trunk mode, (c) Damped branch mode.
	(d) A double-branched Y-shaped structure, and its modes of interest: (e) Trunk mode, (f) Damped large branch mode, (g) Damped small branch mode.
	\label{fig:BIO}}
\end{figure}

Based on the results of Sections~\ref{sec:Model} and~\ref{sec:FEM}, two bioinspired branched structures are considered, figure~\ref{fig:BIO}.

The first bioinspired structure, shown in figure~\ref{fig:T}, is a two-generation T-shaped structure designed so that $\phi_{\rm b} = \pi/2$ at each branching.
The ratios of branch length and diameter are respectively the same between orders of branching, i.e. ${l_2}/{l_1} = {l_3}/{l_2}$ and ${d_2}/{d_1}={d_3}/{d_2}$.
They are chosen so that the modal frequency ratio between the trunk mode, figure~\ref{fig:T_1}, and the last-order branch mode, figure~\ref{fig:T_2}, is $1$:$2$ and so that the modal mass ratio is $0.2$.
A damping rate of $0.2$ is introduced in this last-order branch mode only.
The second bioinspired structure, shown in figure~\ref{fig:S}, consists of a double Y-shaped pattern with an added level of branching at $3/4$ of the height of the trunk.
Both levels of branching have a branching angle $\phi_{\rm b} = 2 \pi/3$ and are designed so that the modal frequency ratio between the trunk mode, figure~\ref{fig:S_1}, and the large branch mode, figure~\ref{fig:S_2}, is $2$, and the modal frequency ratio between the trunk mode and the small branch mode, figure~\ref{fig:S_3}, is $3$.
A damping of $0.2$ is introduced in the two branch modes only.

\begin{figure}[!ht]
	\centering
	\psfrag{x}[b][b]{$E_0$}
	\psfrag{y}[l][l]{$\xif$}
		\psfrag{a1}[bl][bl][0.8]{$.10^{-2}$}
				\psfrag{x0}[l][][0.8]{$0$}
				\psfrag{x1}[][][0.8]{$0.5$}
				\psfrag{x2}[][][0.8]{$1$}
				\psfrag{y0}[br][r][0.8]{$0$}
				\psfrag{y1}[r][r][0.8]{$1$}
				\psfrag{y2}[r][r][0.8]{$2$}
				\psfrag{y3}[r][r][0.8]{$3$}
				\psfrag{y4}[r][r][0.8]{$4$}
	\includegraphics[scale=0.8]{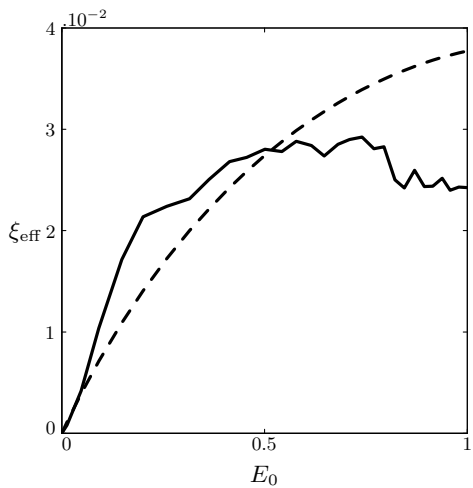}
	\caption{Effect of the initial energy level, $E_0$, on the effective damping, $\xif$, of the two bioinspired structures. (---) Ramified T-shaped structure; (-~-~-) Double-branched Y-shaped structure.
  \label{fig:FEM2_Zeff_vs_E0}}
\end{figure}

The motion is computed using the same finite-element method with the same initial condition, and the same normalization for the energy, as in Section~\ref{sec:FEM}.
The resulting effective damping is studied with the same damping criterion as in Section~\ref{sect:crit} and is plotted in figure~\ref{fig:FEM2_Zeff_vs_E0} as a function of the normalized initial energy. 
In both structures, the effective damping reaches several percent ($\approx 3 \%$) for high levels of initial energy, roughly corresponding to a third of the initial energy being dissipated after one period of the first mode.
These results on two different bioinspired branched structures show that the damping-by-branching mechanism seems to be robust regarding the branching scheme.

\section{Discussion and conclusion} \label{sec:discuss}

At this stage, one may consider the results of the preceding sections in relation to the proposed aim of the paper: to identify and characterize the elementary mechanism that causes nonlinear modal energy transfer and damping in branched structures.
In~Section~\ref{sec:Model}, we have shown that branching is the key ingredient needed to obtain the modal energy transfer and the resulting effective damping that several authors had suspected.
Sections \ref{sec:FEM} and \ref{sec:bio} confirm that the essential features of this damping-by-branching are present even in more complicated branched models.
Since this nonlinear mechanism originates in geometrical effects, the larger the amplitude of motion, the higher the damping.
Efficient use of this mechanism can be made with modest requirements of the design parameters: a modal frequency ratio between the trunk mode and the branch mode near $1$:$2$, a branch damping rate larger than $0.2$, and the highest possible modal mass ratio.
All these results suggest that modal energy transfer and the resulting damping-by-branching is a robust effect at large amplitudes of motion: it may therefore be used in a bioinspired perspective for efficiently damped slender structures.
Clearly, the mechanism found here is distinct from the classical tuned-mass damper, hypothesized by \citeasnoun{Spatz2007} to exist in trees. The present damping-by-branching mechanism differs in two essential points: (i) it is strongly amplitude-dependent; (ii) it is not associated with the condition of identical modal frequencies of the trunk mode and of the branch mode.

Before generalizing our results, discussion is needed of some of the assumptions made to derive them.
Firstly, all the analyses pertaining to the effective damping have been made on the dynamical responses to pull-and-release initial loading of the undamped trunk modes.
This choice was made so that only the nonlinear geometrical effects could cause the effective damping of the structure.
If other classical types of loading were considered such as an initial impulse, or a harmonic or random forcing~\cite{Humar2002}, energy would have been given to all the modes of the branched structure.
Although this would make the global energy balance more complex to analyze, the nonlinear geometric terms responsible for the modal energy transfer~\ref{eq:eqadim} would be still present but the resulting effective damping would not be simply quantified.
Secondly, we have always considered perfectly symmetric and plane structures. 
If asymmetry between the branches is introduced in the model of Section~\ref{sec:Model}, a linear coupling is introduced between the trunk and branch angles of motion so that the energy balance analysis is again more complex.
Similarly, if three-dimensional effects are introduced, such as torsion or multiple 3D branching as in real trees, a much larger number of degrees of freedom is needed in order to describe the dynamics of the structure.
However, the results of Sections \ref{sec:FEM} and \ref{sec:bio} show that complicating the modal content of the model does not impact the existence of the modal transfer mechanism and the resulting effective damping.
Moreover, \citeasnoun{Rodriguez2008} showed that there exist no significant differences between the dynamics of an actual tree architecture and that of an idealized one.

More generally, the question of how branched systems move in a fluid environment is important in practice.
In fact, our analysis on damping originated in the dynamics of trees under wind-loading~\cite{Spatz2007}.
In the interaction between a branched structure and flow, several distinct effects may be expected~\cite{blevins1990,Paidoussis2011}.
Firstly, even in the absence of flow, just the presence of a fluid around the structure causes damping.
This damping is present in all modes, is amplitude-dependent, and introduces nonlinear coupling between modes.
A flow-induced damping may also appear in addition. These effects may be gathered under the generic term of aeroelastic or hydroelastic damping.
Secondly, added mass and added stiffness effects appear and may alter the essential dynamical characteristics of the branched structure, such as frequencies and modal shapes.
These effects are more pronounced in water.
Finally, flow may cause a large variety of loadings through mechanism such as turbulence excitation or wake interactions.
From this list, it appears that all the key parameters involved in the mechanism of damping-by-branching are affected by a fluid environment: modal dampings, frequencies, modal shapes, and external excitations.
A systematic analysis of these effects is clearly needed on the basis of the simple framework presented in this article.

In the design of a slender structure, a bioinspired damping effect can easily be obtained according to the present results.
The main design rule to follow is to introduce branching.
Secondary rules, aiming at optimizing the efficiency of this damping mechanism, are to set ratio between the modal frequency of the branch mode and the trunk mode near $1$:$2$ and the modal mass ratio as high as possible.
Although a design rule based on the ratio of modal frequencies and masses is not common, it should be noted that the requirements are not strict, we have shown in Sections \ref{sec:Model} and \ref{sec:FEM} that damping-by-branching is robust and is significant for a wide range of modal frequency and mass ratios.
The $1$:$2$ rule is only indicative, as it is related to the original internal resonance condition between the branch and trunk modes. 
At this point, the concept of damping-by-branching has only been demonstrated to exist theoretically and numerically. Although it has been inspired by the observation of natural systems, it evidently needs to be explored experimentally on man-made structures.

\section*{Acknowledgements}

The authors gratefully acknowledge the help of Chris Bertram, from the University of Sydney, and Cyril Touz\'e, from ENSTA - UME, for stimulating discussions and useful corrections on the manuscript.
This research was funded by the French Ministry of Defence -- DGA (D\'el\'egation G\'en\'erale pour l'Armement) -- through the Ph.D scholarship program and the contract $2009.60.034.00.470.75.11$.

\bibliographystyle{dcu}

\end{document}